\documentclass[10pt]{article}
\usepackage{amssymb}
\usepackage{amsmath}
\usepackage{graphicx}
\usepackage{fancyhdr}
\usepackage{cite}

\begin{document}

\newcommand{\bin}[2]{\left(\begin{array}{c}\!#1\!\\\!#2\!\end{array}\right)}

\newcommand{\threej}[6]{\left(\begin{array}{ccc} #1 & #2 & #3 \\ #4 & #5 & #6\end{array}\right)}

\huge

\begin{center}
Expression of relativistic expectation values of powers of $r$ in terms of Clebsch-Gordan coefficients
\end{center}

\vspace{0.5cm}

\large

\begin{center}
Jean-Christophe Pain\footnote{jean-christophe.pain@cea.fr}
\end{center}

\normalsize

\begin{center}
CEA, DAM, DIF, F-91297 Arpajon, France
\end{center}

\vspace{0.5cm}

\begin{abstract}
It is shown that, in the relativistic hydrogenic approximation, expectation values of the type $\langle n\ell j|r^k|n\ell j\rangle$ can be expressed in terms of Clebsch-Gordan coefficients or $3jm$ symbols. This generalizes the results obtained by Varshalovich and Karpova [Opt. Spectrosc. {\bf 118}, 1-5 (2015); Opt. Spektrosk. {\bf 118}, 3-7] in the non-relativistic hydrogenic case.
\end{abstract}

\section{Introduction}

The apparatus of the quantum theory of angular momentum is widely used to calculate the angular parts of matrix elements in various quantum-mechanical problems. The reason is that the corresponding special functions (the spherical functions, the $D$ functions, the Wigner $3jm$ symbols, and so forth) have been studied in detail and tabulated \cite{varshalovich88}. We use the notation of the Clebsch-Gordan coefficient of Varshalovich \emph{et al.} of the latter reference, which was used by Jahn \cite{jahn51} and Alder \cite{alder52}:

\begin{equation}\label{cle}
C_{j_1m_1,~j_2m_2}^{jm}=(-1)^{j_1-j_2+m}\sqrt{2j+1}\threej{j_1}{j_2}{j}{m_1}{m_2}{-m}.
\end{equation}

\noindent The other notations (Wigner \cite{wigner59,wang70}, Eckart \cite{eckart30}, Van der Waerden \cite{waerden32}, Condon and Shortley \cite{condon35},  Boys \cite{boys51}, Blatt and Weisskopf \cite{blatt52}, Biedenharn \cite{biedenharn52}, Rose \cite{rose54} and Fano \cite{fano52}) are given at p. 52 of Ref. \cite{edmonds}. Several expressions of Clebsch-Gordan coefficients in terms of hypergeometric $~_3F_2$ functions were published by different authors \cite{smorodinskii72}, such as Fock \cite{fock68}, Bandzaitis and Yutsis \cite{bandzaitis}, Majumdar \cite{majumdar55} or Racah \cite{racah42}. In 1973, Karasev and Shelepin pointed out the interest of an intimate relation between the calculation of finite differences, hypergeometric series and the theory of Clebsch-Gordan coefficients \cite{karasev73}. In 1979, Varshalovich and Khersonskii found a simple connection between $\langle r^k\rangle=\langle n\ell|r^k|n\ell\rangle$ and $C_{\ell~n,(k+1)~0}^{\ell~n}$ \cite{varshalovich79}. In 2015, Varshalovich and Karpova found that both $C_{\ell'~n,(k+1)~0}^{\ell~n}$ and $\langle n\ell|r^k|n\ell'\rangle$ can be put in a form in which they are proportional to the hypergeometric function \cite{luke69}:

\begin{equation}
_3F_2\left(\begin{array}{l}
\ell+\ell'-k,\ell-\ell'-k-1,-k-1\\
n+\ell-k,-2k-2\\
\end{array};1\right).
\end{equation}

They obtained the following result \cite{varshalovich15}:

\begin{equation}
\frac{\langle n\ell|r^k|n\ell'\rangle}{C_{\ell'~n,(k+1)~0}^{\ell~n}}=\frac{i^{\Delta}}{2n\sqrt{2\ell+1}}\left(\frac{n}{2}\right)^kf_{\ell,\ell'}^k,
\end{equation}

\noindent where

\begin{equation}
f_{\ell,\ell'}^k=\left[\frac{(k+1-\Delta)!(k+1+\Delta)!(\ell+\ell'+k+2)!}{\left[(k+1)!\right]^2(\ell+\ell'-k-1)!}\right]^{1/2},
\end{equation}

\noindent with $\Delta=\ell-\ell'\geq$ 0, and the Clebsch-Gordan $C_{\ell'~n,(k+1)~0}^{\ell~n}$ is evaluated in a nonphysical region of its arguments, where the so-called triangle condition is not satisfied and the projections of angular momenta $\ell$ and $\ell'$ are larger than the angular momenta themselves. In this region, the Clebsch-Gordan coefficient vanishes, but the product $f_{\ell,\ell'}^k\times C_{\ell'~n,(k+1)~0}^{\ell~n}$ is nonzero. For $(k+1)<0$, coefficient $C_{\ell'~n,(k+1)~0}^{\ell~n}$ transforms into $(-1)^{\ell-\ell'}C_{\ell'~n,-(k+2)~0}^{\ell~n}$ \cite{bandzaitis64} and $f_{\ell,\ell'}^k$ into $f_{\ell,\ell'}^{-k}$. For negative values, one has $(-a)!/(-b)!=(-1)^{b-a}(b+1)!/(a+1)!$.

All these features for the radial aspect of the hydrogen atom may seem surprising, given no obvious rotational symmetry considerations for the radial equation. However, they are understood through the recognition that the radial problem has the symmetry of the non-compact group O(2,1) \cite{judd70,armstrong71,heim09}. This group's closed triplet of operators under commutation is very similar, apart from sign changes in the structure factors, to those of the angular momentum's O(3) triplet, which explains the appearance of such Clebsch-Gordan coefficients in radial matrix elements.

In the present work, we propose to follow the philosophy of the works of Varshalovich, Khersonskii and Karpova \cite{varshalovich79,varshalovich15} in order to obtain an expression of $\langle r^k\rangle=\langle n\ell j|r^k|n\ell j\rangle$ in the relativistic case. Instead of $|n\ell j\rangle$, the notation $|n\kappa\rangle$ with $\kappa=(\ell-j)(2j+1)$ is often used. One has $\kappa=\mp\left(j+1/2\right)$ for $j=\ell\pm 1/2$, i.e. $\kappa=-\ell-1$ for $j=\ell+1/2$ and $\kappa=\ell$ for $j=\ell-1/2$.

\section{Expectation values of $r^k$ in the relativistic case}

The expectation values of $r^k$ in the relativistic case read

\begin{equation}
\langle r^k\rangle=\int_0^{\infty}r^k\left[P_{n\kappa}(r)^2+Q_{n\kappa}^2(r)\right]dr,
\end{equation}

\noindent where the normalized bound-state solutions $P_{n\kappa}(r)$ and $Q_{n\kappa}(r)$ of the Dirac equation are respectively

\begin{equation}
P_{n\kappa}(r)=\mathcal{N}_{n\kappa}\sqrt{1+\frac{n_r+\gamma}{N}}x^{\gamma}e^{-x/2}\left\{(N-\kappa)~_1F_1\left(\begin{array}{c}
-n_r\\
2\gamma+1
\end{array}; x\right)-n_r~_1F_1\left(\begin{array}{c}
-n_r+1\\
2\gamma+1
\end{array}; x\right)\right\}
\end{equation}

\noindent and

\begin{equation}
Q_{n\kappa}(r)=-\mathcal{N}_{n\kappa}\sqrt{1-\frac{n_r+\gamma}{N}}x^{\gamma}e^{-x/2}\left\{(N-\kappa)~_1F_1\left(\begin{array}{c}
-n_r\\
2\gamma+1
\end{array}; x\right)+n_r~_1F_1\left(\begin{array}{c}
-n_r+1\\
2\gamma+1
\end{array}; x\right)\right\},
\end{equation}

\noindent with the normalization constant

\begin{equation}
\mathcal{N}_{n\kappa}=\frac{1}{N}\sqrt{\frac{Z}{2(N-\kappa)}\frac{\mathrm{\Gamma}(n_r+2\gamma+1)}{\left[\mathrm{\Gamma}(2\gamma+1)\right]^2n_r!}}.
\end{equation}

\noindent One has $\zeta=2Z/N$, $x=\zeta r$, $\gamma=\sqrt{\kappa^2-\alpha^2Z^2}$ ($\alpha$ being the fine-structure constant), $n_r=n-|\kappa|$, $N=\sqrt{n_r^2+2n_r\gamma+\kappa^2}$ and $~_1F_1$ represents the confluent (Kummer) hypergeometric function. In general, the computation of $\langle r^k\rangle$ in the relativistic case is more complicated than in the non-relativistic case. This is probably the reason why explicit expressions are not widely known. The first explicit expression was given by Davies \cite{davies39}.  
For a long time, values were available only for $k=-3,-1,1$ and $2$ from the work of Garstang and Mayers \cite{garstang66} and Burke and Grant \cite{burke67}. Various methods have been applied to obtain these expectation values (see for instance Refs. \cite{crubellier71,johnson79,goldman82,salamin95a,salamin95b}). The topic was also extended to non-diagonal matrix elements and recurrence relations are known for basic integrals and other closely related quantities (see the non-exhaustive list of references \cite{epstein62,bessis85,kobus87,shabaev91,andrae97,martinez00,martinez01,dong04,dong05}). The functions $P_{n\kappa}(r)$ and $Q_{n\kappa}(r)$ can be expressed in terms of generalized Laguerre polynomials, and the determination of diagonal (expectation values) and non-diagonal matrix elements boils down to the calculation of integrals of the kind \cite{szmytkowski97}:

\begin{equation}\label{ii}
I_{mn}^{\alpha\beta\gamma}(a,b,c)=\int_0^{\infty}x^{\gamma}e^{-cx}L_m^{(\alpha)}(ax)L_n^{(\beta)}(bx)dx,
\end{equation}

\noindent where $L_p^{(q)}$ are generalized Laguerre polynomials \cite{sakurai93}. Integral (\ref{ii})  can be calculated using \cite{pain19}

\begin{equation}
I_{mn}^{\alpha\beta\gamma}(a,b,c)=\frac{1}{m!n!}\frac{\partial^m}{\partial s^m}\frac{\partial^n}{\partial t^n}\frac{\Gamma(\gamma+1)}{(1-s)^{\alpha+1}(1-t)^{\beta+1}\left(c+a\frac{s}{1-s}+b\frac{t}{1-t}\right)^{\gamma+1}},
\end{equation}

\noindent which can be obtained using the generating-function method \cite{bransden83}. Mart\'inez-y-Romero \emph{et al.} \cite{martinez00,martinez01} obtained recurrence formulae by an approach inspired by the relativistic extension of the second  hypervirial  method  that  has  been  successfully  employed  to  deduce  an  analogous relation in non-relativistic quantum mechanics. The latter authors used this relation to deduce relativistic versions of the Pasternack-Sternheimer rule \cite{pasternack37a,pasternack37b,pasternack62} and of the virial theorem. The connection of the expectation values with the Chebyshev and Hahn polynomials of a discrete variable was studied by Suslov \cite{suslov09}, who derived two sets of Pasternack-type matrix identities for these integrals \cite{suslov10,suslov10b}. Suslov also studied integrals of the type

\begin{equation}
\int_0^{\infty}r^k\left[P_{n\kappa}(r)^2-Q_{n\kappa}^2(r)\right]dr,
\end{equation}

\noindent and

\begin{equation}
\int_0^{\infty}r^kP_{n\kappa}(r)Q_{n\kappa}(r)dr,
\end{equation}

\noindent which we do not consider here. The complexity of the expressions of $\langle r^k\rangle$ increases with $k$ (see the examples of $\langle r^{-3}\rangle$, $\langle r^{-2}\rangle$, $\langle r^{-1}\rangle$, $\langle r\rangle$ and $\langle r^2\rangle$ in table \ref{tab1}). In the relativistic case, we have \cite{andrae97}:

\begin{eqnarray}\label{eq1}
\langle r^k\rangle&=&\frac{1}{2N}\frac{1}{\zeta^k}\frac{\Gamma(2\gamma+n_r+k)}{\Gamma(2\gamma+n_r+1)}\left\{(N-\kappa)(2\gamma+n_r+k)_3F_2\left(\begin{array}{l}
-k,-k,-n_r\\
1,-2\gamma-n_r-k\\
\end{array};1\right)\right.\nonumber\\
& &+(N+\kappa)(2\gamma+n_r)_3F_2\left(\begin{array}{l}
-k,-k,-n_r+1\\
1,-2\gamma-n_r-k+1\\
\end{array};1\right)\nonumber\\
& &+\left.\frac{2k}{N}n_r(n_r+\gamma)(n_r+2\gamma)~_3F_2\left(\begin{array}{l}
-k,-k+1,-n_r+1\\
2,-2\gamma-n_r-k+1\\
\end{array};1\right)\right\},
\end{eqnarray}

\noindent where $\Gamma$ is the usual Gamma function. The formula (21) of Varshalovich, which is also the van der Waerden form \cite{waerden32,smorodinskii72} gives:

\begin{eqnarray}
_3F_2\left(\begin{array}{l}
a,b,c\\
d,e\\
\end{array};1\right)&=&\sqrt{\frac{\Gamma(1-a)\Gamma(1-b)\Gamma(1-c)}{(d+e-b-c-1)\Gamma(d-a)\Gamma(d-b)\Gamma(d-c)}}\nonumber\\
& &\times\sqrt{\frac{\Gamma(-a-b-c+d+e)}{\Gamma(e-a)\Gamma(e-b)\Gamma(e-c)}}\Gamma(d)\Gamma(e)\Lambda(a,b,c,d,e),
\end{eqnarray}

\noindent where

\begin{equation}
\Lambda(a,b,c,d,e)=C_{\frac{d-a-b-1}{2}~\frac{b+d-a-1}{2},\frac{e-a-c-1}{2}~\frac{a-c-e+1}{2}}^{\frac{d+e-b-c}{2}-1~\frac{b+d-c-e}{2}},
\end{equation}

\noindent $C$ being the Clebsch-Gordan coefficient defined by Eq. (\ref{cle}).

\begin{table}[ht]
\centering
\begin{tabular}{|c|c|}\hline\hline
$\;\;\;\;\;\;\; k \;\;\;\;\;\;\;$ & $\langle r^k \rangle$ \\\hline\hline
-3 & $\frac{8Z^3\left[6\kappa^2\left(n_r+\gamma\right)^2+2N^2\left(1-\gamma^2\right)-6N\kappa\left(n_r+\gamma\right)\right]}{(2\gamma-2)(2\gamma-1)2\gamma(2\gamma+1)(2\gamma+2)N^5}$ \\
   & = $\frac{2Z^3\left[-3N^2\kappa\sqrt{1-Z^2\alpha^2/N^2}+N^2+2\gamma^2\kappa^2+\left(N^2-\kappa^2\right)\left(3\kappa^2-\gamma^2\right)\right]}{(\gamma-1)\gamma(\gamma+1)(2\gamma-1)(2\gamma+1)N^5}$ \\
-2 & $\frac{4Z^2\left[2\kappa^2\left(n_r+\gamma\right)-N\kappa\right]}{(2\gamma-1)2\gamma(2\gamma+1)N^4}$ \\
-1 & $\frac{Z(\kappa^2+n_r\gamma)}{\gamma N^3}$ \\
 1 & $\frac{3N^2\left(n_r+\gamma\right)-N\kappa-\kappa^2\left(n_r+\gamma\right)}{2ZN}$ \\
   & = $\frac{\left(3N^2-\kappa^2\right)\sqrt{1-Z^2\alpha^2/N^2}-\kappa}{2Z}$\\
 2 & $\frac{N^2\left[5\left(n_r+\gamma\right)^2+1-\gamma^2\right]-6N\kappa\left(n_r+\gamma\right)-4\kappa^2\left(n_r+\gamma\right)^2}{2Z^2}$ \\
   & = $\frac{N^2\left\{\left(5N^2-2\kappa^2\right)\left(1-Z^2\alpha^2/N^2\right)+\left(1-\gamma^2\right)-3\kappa \sqrt{1-Z^2\alpha^2/N^2}\right\}}{2Z^2}$\\\hline\hline
\end{tabular}
\caption{Expressions of $\langle r^k\rangle$ for $k$=-3, -2, -1, 1 and 2 in the relativistic (Dirac) case.}\label{tab1}
\end{table}

\noindent The notation $\Gamma(-m)$, where $m$ is a positive integer, means $\Gamma(-m)=(-1)^{m-1}(m-1)!$. One has therefore the three relations

\begin{eqnarray}
_3F_2\left(\begin{array}{l}
-k,-k,-n_r\\
1,-2\gamma-n_r-k\\
\end{array};1\right)
&=&(-1)^{1+k+n_r+2\gamma}\sqrt{\frac{\Gamma(1+k-2\gamma)}{\Gamma(-k-2\gamma)}}\frac{\Gamma(-k-n_r-2\gamma)}{\Gamma(-n_r-2\gamma)}\nonumber\\
& &\times\threej{k}{-\frac{1}{2}-\gamma}{-\frac{1}{2}-\gamma}{0}{\frac{1}{2}+n_r+\gamma}{-\frac{1}{2}-n_r-\gamma},
\end{eqnarray}

\begin{eqnarray}
_3F_2\left(\begin{array}{l}
-k,-k,-n_r+1\\
1,-2\gamma-n_r-k+1\\
\end{array};1\right)&=&(-1)^{k+n_r+2\gamma}\sqrt{\frac{\Gamma(1+k-2\gamma)}{\Gamma(-k-2\gamma)}}\frac{\Gamma(1-k-n_r-2\gamma)}{\Gamma(1-n_r-2\gamma)}\nonumber\\
& &\times\threej{k}{-\frac{1}{2}-\gamma}{-\frac{1}{2}-\gamma}{0}{-\frac{1}{2}+n_r+\gamma}{\frac{1}{2}-n_r-\gamma}
\end{eqnarray}

\noindent and

\begin{eqnarray}
_3F_2\left(\begin{array}{l}
-k,-k+1,-n_r+1\\
2,-2\gamma-n_r-k+1\\
\end{array};1\right)
&=&\frac{(-1)^{1+k+n_r+2}\Gamma(1-k-n_r-2\gamma)}{\sqrt{k(k+1)}}\nonumber\\
& &\times\sqrt{\frac{\Gamma(1+k-2\gamma)}{n_r\Gamma(-k-2\gamma)\Gamma(-n_r-2\gamma)\Gamma(1-n_r-2\gamma)}}\nonumber\\
& &\times\threej{k}{-\frac{1}{2}-\gamma}{-\frac{1}{2}-\gamma}{1}{-\frac{1}{2}+n_r+\gamma}{-\frac{1}{2}-n_r-\gamma},
\end{eqnarray}

\noindent and the final result is then

\begin{eqnarray}\label{eq2}
\langle r^k\rangle&=&\frac{1}{2N}\frac{1}{\zeta^k}\frac{\Gamma(2\gamma+n_r+k)}{\Gamma(2\gamma+n_r+1)}\nonumber\\
& &\times\left\{(N-\kappa)(2\gamma+n_r+k)(-1)^{1+k+n_r+2\gamma}\sqrt{\frac{\Gamma(1+k-2\gamma)}{\Gamma(-k-2\gamma)}}\frac{\Gamma(-k-n_r-2\gamma)}{\Gamma(-n_r-2\gamma)}\right.\nonumber\\
& &\times\threej{k}{-\frac{1}{2}-\gamma}{-\frac{1}{2}-\gamma}{0}{\frac{1}{2}+n_r+\gamma}{-\frac{1}{2}-n_r-\gamma}\nonumber\\
& &+(N+\kappa)(2\gamma+n_r)(-1)^{k+n_r+2\gamma}\sqrt{\frac{\Gamma(1+k-2\gamma)}{\Gamma(-k-2\gamma)}}\frac{\Gamma(1-k-n_r-2\gamma)}{\Gamma(1-n_r-2\gamma)}\nonumber\\
& &\times\threej{k}{-\frac{1}{2}-\gamma}{-\frac{1}{2}-\gamma}{0}{-\frac{1}{2}+n_r+\gamma}{\frac{1}{2}-n_r-\gamma}\nonumber\\
& &+\frac{2k}{N}n_r(n_r+\gamma)(n_r+2\gamma)(-1)^{1+k+n_r+2\gamma}\Gamma(1-k-n_r-2\gamma)\nonumber\\
& &\times\sqrt{\frac{\Gamma(1+k-2\gamma)}{k(k+1)n_r\Gamma(-k-2\gamma)\Gamma(-n_r-2\gamma)\Gamma(1-n_r-2\gamma)}}\nonumber\\
& &\left.\times\threej{k}{-\frac{1}{2}-\gamma}{-\frac{1}{2}-\gamma}{1}{-\frac{1}{2}+n_r+\gamma}{-\frac{1}{2}-n_r-\gamma}\right\}.
\end{eqnarray}

As in the non-relativistic case, some $3j$ coefficients are evaluated in a nonphysical region of their arguments and vanish for that reason; however, the products of themselves by their prefactors are finite. It is also possible to derive other expressions of $\langle r^k\rangle$ in terms of Clebsch-Gordan coefficients, for instance using the additional relation \cite{andrae97}:

\begin{eqnarray}\label{eq3}
\langle r^k\rangle&=&\frac{1}{2N}\frac{1}{\zeta^k}\frac{\Gamma(2\gamma+n_r+k)}{\Gamma(2\gamma+n_r+1)}\frac{1}{(4\gamma+2n_r+k+1)}\nonumber\\
& &\times\left\{\vphantom{ _3F_2\left(\begin{array}{l}
-k,-k,-n_r+1\\
1,-2\gamma-n_r-k+1\\
\end{array};1\right)}-[(4\gamma+2n_r+k+1)\kappa(N-\kappa)+(n_r+2\kappa)^2(2\gamma+k+1)]\right.\nonumber\\
& &\times (2\gamma+n_r+k) ~_3F_2\left(\begin{array}{l}
-k,-k,-n_r\\
1,-2\gamma-n_r-k\\
\end{array};1\right)\nonumber\\
& &+2(n_r+\gamma)(n_r+2\gamma)(2\gamma+n_r+k)(2\gamma+n_r+k+1)~_3F_2\left(\begin{array}{l}
-k-1,-k-1,-n_r\\
1,-2\gamma-n_r-k-1\\
\end{array};1\right)\nonumber\\
& &+\left.(N+\kappa)(2\gamma+n_r)\left[N(2\gamma+k+1)+2\kappa(n_r+\gamma)\right]~_3F_2\left(\begin{array}{l}
-k,-k,-n_r+1\\
1,-2\gamma-n_r-k+1\\
\end{array};1\right)\right\},\nonumber\\
\end{eqnarray}

\noindent which will yield another expression of $\langle r^k\rangle$. Although the coefficients in front of the hypergeometric $_3F_2$ functions are more complicated than those in Eq. (\ref{eq1}), the latter equation has an advantage for the recursion with respect to $k$. A recursion based on such equation requires only two, instead of three, sequences of the $_3F_2$ series, which, in addition, are related to each other in a simpler way than in Eq. (\ref{eq1}). It is interesting also to mention, for checking purposes for instance, that in the case where $n_r=0$, $\langle r^k\rangle$ reduces to

\begin{equation}
\langle r^k\rangle=\frac{(N-\kappa)}{2N}\frac{1}{\zeta^k}\frac{\Gamma(2\gamma+k+1)}{\Gamma(2\gamma+1)}=\frac{(2\gamma+1)_k}{\zeta^k},
\end{equation}

\noindent where $(a)_n=a(a+1)(a+2)\cdots(a+n-1)$ represents the Pochhammer symbol.

\section{Conclusion}

Varshalovich and Karpova found that, in the non-relativistic (Schr\"odinger) case, radial integrals $\langle n\ell|r^k|n\ell'\rangle$ can be expressed in terms of Clebsch-Gordan coefficients. We have shown that this is also the case for expectation values in the relativistic case, and we have provided the corresponding relations. In the future, we plan to investigate the case of relativistic non-diagonal elements.

\end{document}